\documentclass[11pt,a4paper]{article}
\usepackage{jheppub}

\usepackage{amsfonts}
\usepackage{slashed}

\def\tr{{\rm Tr}}

\def\bea{\begin{eqnarray}}
\def\eea{\end{eqnarray}}
\def\nn{\nonumber}
\def\half{\frac{1}{2}}

\def\lmatrix{\left(\begin{array}}
\def\rmatrix{\end{array}\right)}

\def\msbar{\overline{\rm MS\kern-0.5pt}\kern0.5pt}
\def\sss{{Symanzik-Symanzik-Symanzik} }
\def\ssc{{Symanzik-Symanzik-clover} }

\def\phat{{\hat p}}
\def\ptilde{{\tilde p}}

\title{The lattice gradient flow at tree-level and its improvement}

\author[abc]{Zoltan Fodor,}
\author[de]{Kieran Holland,}
\author[f]{Julius Kuti,}
\author[c]{Santanu Mondal,}
\author[c]{Daniel Nogradi,}
\author[f]{and Chik Him Wong}

\affiliation[a]{University of Wuppertal, Department of Physics, Wuppertal D-42097, Germany}
\affiliation[b]{J\"ulich Supercomputing Center, Forschungszentrum J\"ulich, J\"ulich D-52425, Germany}
\affiliation[c]{E\"otv\"os University, Institute for Theoretical Physics, Budapest 1117, Hungary}
\affiliation[d]{University of the Pacific, 3601 Pacific Ave, Stockton CA 95211, USA}
\affiliation[e]{Albert Einstein Center for Fundamental Physics, Institute
for Theoretical Physics, Bern University, Sidlerstrasse 5, CH-3012 Bern, Switzerland}
\affiliation[f]{University of California, San Diego, 9500 Gilman Drive, La Jolla, CA 92093, USA}

\emailAdd{fodor@bodri.elte.hu}
\emailAdd{kholland@pacific.edu}
\emailAdd{jkuti@ucsd.edu}
\emailAdd{santanu@bodri.elte.hu}
\emailAdd{nogradi@bodri.elte.hu}
\emailAdd{rickywong@physics.ucsd.edu}

\abstract{ 
The Yang-Mills gradient flow and the observable $\langle E(t) \rangle$, defined by the square of the field strength
tensor at $t>0$, are calculated at 
finite lattice spacing and tree-level in the gauge coupling. 
Improvement of the flow, the gauge action and the observable are all considered. 
The results are relevant for two purposes. First, the discretization of the flow, gauge action and
observable can be chosen in such a way that $O(a^2)$, $O(a^4)$ or even $O(a^6)$ improvement is achieved.
Second, simulation results
using arbitrary discretizations can be tree-level improved by the perturbatively calculated correction factor normalized
to one in the continuum limit.
}


\begin{document}

\maketitle

\section{Introduction and summary}
\label{intro}

The lattice version of the Yang-Mills gradient flow \cite{Luscher:2009eq, Luscher:2010iy, Luscher:2010we, Luscher:2011bx}
has been proven to be extremely useful in simulations of lattice gauge theories. Applications include scale setting
\cite{Borsanyi:2012zs, Bazavov:2013gca, Sommer:2014mea},
measuring the renormalized gauge coupling \cite{Fodor:2012td,Fodor:2012qh, Fritzsch:2013je, Fritzsch:2013hda,
Ramos:2013gda, Fritzsch:2013yxa, Rantaharju:2013bva, Luscher:2014kea} and
thermodynamics \cite{Asakawa:2013laa}, while some of the important theoretical
developments include chiral perturbation theory aspects \cite{Bar:2013ora}, its relation to the energy momentum 
tensor \cite{Suzuki:2013gza, DelDebbio:2013zaa, Makino:2014taa} and chiral symmetry \cite{Luscher:2013cpa, Shindler:2013bia}.

In all of these applications the size of cut-off effects is an important question and so far has not been systematically
explored, although see \cite{Cheng:2014jba} for some work in this direction. 
The observable most commonly used is $E(t)$, the expectation value of the field strength squared at $t>0$. Its
discretization is affected by three building blocks of the calculation: 
(1) a choice needs to be made for the gauge action used along the gradient flow,
(2) one needs to choose a discretized dynamical gauge action used for generating the
configurations, and finally (3) one needs
to choose a discretization for the observable $E$. Notice that in the continuum all three choices define the
same 
$F_{\mu\nu} F_{\mu\nu}$ corresponding to three (potentially different)
discretizations of $F_{\mu\nu} F_{\mu\nu}$.

In this work a broad family of discretizations is considered. Each of the three building blocks is chosen to be the
Symanzik improved gauge action with three different improvement coefficients,
$c_{1f}$ for the flow, $c_{1g}$ for the dynamical action and $c_{1e}$ for the observable. For the
observable $E$ we will also consider the symmetric clover-type discretization which does not depend on any parameters. 

The goal of the present work is twofold. First, we would like to calculate the optimal values for the parameters
$c_{1f}$, $c_{1g}$ and $c_{1e}$ in order to achieve as high an order of improvement as possible. We will see that
it is possible to choose these 3 parameters such that $O(a^6)$ improvement is achieved. If the clover observable is used
$c_{1f}$ and $c_{1g}$ can be chosen to achieve $O(a^4)$ improvement. If the dynamical gauge action is considered fixed,
i.e. $c_{1g}$ is considered fixed,
$c_{1f}$ and $c_{1e}$ can be chosen to arrive at $O(a^4)$ improvement, or if the clover observable is used then $c_{1f}$
may be chosen to achieve $O(a^2)$ improvement.

Second, if simulations and measurements are performed with any set of parameters the resulting $\langle t^2 E(t)
\rangle$ expectation value may be tree-level improved by the perturbatively calculated correction factor normalized to
one in the continuum limit. The improved data will lead to the same
continuum limit as the unimproved one but the size of cut-off effects will be smaller. We will discuss the calculation
of the correction factor both in finite and infinite volume.

Since our discussion is at tree-level, fermions play no role. In the following the index $1$ will be dropped from the
$c_1$ improvement coefficients and hence they will be labelled as $c_f,c_g,c_e$.

The organization of the paper is as follows. In section \ref{gradientflow} the gradient flow is reviewed briefly, in
section \ref{latpert} the lattice perturbation theory setup at tree-level is outlined and the
results are given up to $O(a^8)$. Various improved flow setups are explained in section \ref{highlyimp}
whereas the application of our formulae for scale setting in QCD is given in section \ref{scaleset}. Finite volume
effects are discussed in section \ref{finitevolume} which are important for running coupling applications. The usefulness of
tree-level improvement is demonstrated numerically in section \ref{numericaltest} using previously published data.
Finally in section \ref{conc} we close with a conclusion and outline several directions to pursue in the future along
the lines presented in this paper.

\section{Gradient flow}
\label{gradientflow}

The gradient flow \cite{Luscher:2009eq,Luscher:2010iy,Luscher:2010we} evolves the gauge field 
$A_\mu$ in an auxiliary variable $t$ by
\bea
\frac{d A_\mu(t)}{dt} = - \frac{\delta S_{\rm YM}}{\delta A_\mu}\;,
\eea
where $S_{\rm YM}$ is the pure Yang-Mills action. Clearly $t$ is of mass dimension $-2$.
Once an observable is given, ${\cal O}(A_\mu)$, the idea of the gradient flow framework is that in the path integral one
integrates over the initial condition $A_\mu(0)$ while the observable is evaluated at $t>0$, $\langle {\cal O}(A_\mu(t))
\rangle$. Under certain circumstances originally $UV$-divergent composite operators become finite for $t>0$ and one may think
of the flow as a renormalization prescription.

In most applications the observable is $E = - \half \tr F_{\mu\nu} F_{\mu\nu}$. 
The resulting finite $t$-dependent quantity can be expanded in renormalized perturbation theory as \cite{Luscher:2010iy}
\bea
\langle t^2 E(t) \rangle = \frac{3(N^2-1)g^2}{128\pi^2}\left( 1 + O(g^2) \right)\;,
\eea
where $g$ is a suitably defined renormalized coupling, for instance in $\msbar$. The corresponding bare perturbative
series is similar with $g$ replaced by $g_0$, the bare coupling. The first term, proportional to $O(g_0^2)$, is a tree
level contribution and the remainder are loop corrections.

Our conventions are identical to \cite{Fodor:2012td,Fodor:2012qh} and for more details about the perturbative aspects of
the gradient flow in a continuum setup see \cite{Luscher:2011bx}.

\section{Lattice perturbation theory at tree-level}
\label{latpert}

Our goal is to compute the lattice spacing dependence of the tree-level term, i.e. we are after the quantity $C(a^2/t)$,
where
\bea
\langle t^2 E(t) \rangle_a = \frac{3(N^2-1)g_0^2}{128\pi^2}\left( C(a^2/t) + O(g_0^2) \right)\;.
\eea
The only scale that can make the quantity $C$ dimensionless is $t$, hence the dependence on $a^2/t$. Let us define the
coefficients $C_{2m}$ by
\bea
\label{cat}
C(a^2/t) = 1 + \sum_{m=1}^\infty C_{2m} \frac{a^{2m}}{t^m}\;.
\eea

In a perturbative calculation \cite{Weisz:1982zw, Symanzik:1983dc, Weisz:1983bn, Luscher:1984xn, Luscher:1985zq} 
one expands in the bare coupling $g_0$ which amounts to an expansion in the gauge field
$A_\mu(x)$. On the lattice the basic variable is the link $U_\mu(x)$ and the two are related by 
\bea
\label{up2}
U_\mu(x) = \exp\left( a g_0 A_\mu(x+ae_\mu/2) \right)\;.
\eea

The gauge field in momentum space will be labelled by $A_\mu(p)$. Let us introduce
\bea
{\hat p}_\mu &=& \frac{2}{a} \sin\left( \frac{ap_\mu}{2} \right) \nn \\
{\tilde p}_\mu &=& \frac{1}{a} \sin( a p_\mu )\;,
\eea
and also $\phat^2 = \sum_\mu \phat_\mu^2$ and $\ptilde^2 = \sum_\mu \ptilde_\mu^2$ for the lattice momenta.

In momentum space the Symanzik improved action is \cite{Weisz:1982zw, Weisz:1983bn, Luscher:1984xn, Luscher:1985zq},
\bea
\label{syms}
S_{\mu\nu} = \delta_{\mu\nu} \left( \phat^2 - a^2 c \sum_\rho \phat_\rho^4 - a^2 c \phat_\mu^2 \phat^2 \right) 
- \phat_\mu \phat_\nu \left( 1 - a^2 c ( \phat_\mu^2 + \phat_\nu^2 ) \right) \;.
\eea
The clover discretization of the field strength tensor on the other hand is \cite{Fritzsch:2013je}
\bea
\label{clov}
K_{\mu\nu} = \left( \delta_{\mu\nu} \ptilde^2 - \ptilde_\mu \ptilde_\nu \right) 
\cos\left( \frac{ap_\mu}{2} \right) \cos\left( \frac{ap_\nu}{2} \right)\;.
\eea

Since we would like to consider a general setup where the discretization for the flow and the gauge action is $S_{\mu\nu}$
with potentially different $c$ improvement coefficients and the observable $E$ can be either $S_{\mu\nu}$ with yet
another improvement coefficient or the clover expression $K_{\mu\nu}$,  let us introduce the general notation 
${\cal S}_{\mu\nu}$ by
\bea
{\cal S}^{f}_{\mu\nu} &=& S_{\mu\nu}(c=c_f) \nn \\
{\cal S}^{g}_{\mu\nu} &=& S_{\mu\nu}(c=c_g)  \\
{\cal S}^{e}_{\mu\nu} &=& S_{\mu\nu}(c=c_e)\;, \qquad {\rm or} \qquad K_{\mu\nu}\;. \nn
\eea
Whether the observable is the clover expression or the Symanzik improved action will be clear from the context.

In terms of the gauge field $A_\mu(p)$ lattice gauge transformations are the usual ones to lowest order,
\bea
A_\mu(p) \to A_\mu(p) - i {\hat p}_\mu\;.
\eea
It is simple to check that both the Symanzik action and the clover observable are transverse, i.e.
\bea
S_{\mu\nu} {\hat p}_\nu &=& 0 \nn \\
E_{\mu\nu} {\hat p}_\nu &=& 0\;.
\eea
It is convenient to add a suitable gauge fixing term to both the propagator and the gradient flow
\bea
{\cal G}_{\mu\nu} = \frac{1}{\alpha} {\hat p}_\mu {\hat p}_\nu\;.
\eea

The continuum flow in section \ref{gradientflow} at finite lattice spacing and tree-level is then
\bea
\frac{d A_\mu(p,t)}{dt} = - \left( {\cal S}^f_{\mu\nu}(p) + {\cal G}_{\mu\nu}(p) \right) A_\nu(p,t)
\eea
which is easy to solve,
\bea
\label{flowsol}
A_\mu(p,t) = \left[ e^{ -t\left( {\cal S}^f + {\cal G} \right) } \right]_{\mu\nu} A_\nu(p,0)
\eea
where on the right hand side we have a matrix exponential. 
Remember that the path integral is over $A_\mu(p,0)$. Our observable is then, at tree-level,
\bea
\langle t^2E(t)\rangle = - \frac{N^2-1}{2}  g_0^2 t^2 \int_{-\frac{\pi}{a}}^{\frac{\pi}{a}} \frac{d^4p}{(2\pi)^4} {\cal S}^e_{\mu\nu}(p) \langle
A_\mu(p,t) A_\nu(-p,t) \rangle\;,
\eea
where the gauge field is now understood as a $U(1)$ field and the color factor $N^2-1$ has already been factored out.
Substituting (\ref{flowsol}) into the above and using the free propagator
\bea
\langle A_\mu(p,0) A_\nu(-p,0) \rangle = - \left[ ({\cal S}^g + {\cal G} )^{-1} \right]_{\mu\nu}
\eea
we obtain
\bea
\label{eee}
\langle t^2E(t) \rangle =\frac{N^2-1}{2}  g_0^2 t^2\int_{-\frac{\pi}{a}}^{\frac{\pi}{a}} \frac{d^4p}{(2\pi)^4} 
\tr\, \left( e^{-t\left({\cal S}^f + {\cal G}\right)} ({\cal S}^g + {\cal G})^{-1} e^{-t\left({\cal S}^f + {\cal
G}\right)} {\cal S}^e \right)\;.
\eea
This expression will be the starting point for all that follows. 

In order to expand in the lattice spacing we simply have to expand ${\cal S}$ which of course involves expanding
$S_{\mu\nu}$ and $K_{\mu\nu}$ as well as the the gauge fixing term ${\cal G}_{\mu\nu}$.

Note that since generally $c_f \neq c_g$ the two exponentials in (\ref{eee}) cannot 
be combined because ${\cal S}^f$ and ${\cal S}^g$
do not commute. The expansions and further calculations are simplest with the choice $\alpha = 1$ but of course the
final result should be $\alpha$-independent. We have checked this for all the final correction coefficients explicitly
and it is easy to see that $\alpha$-independence holds to all orders.

\begin{figure}
\begin{center}
\includegraphics[width=7.5cm]{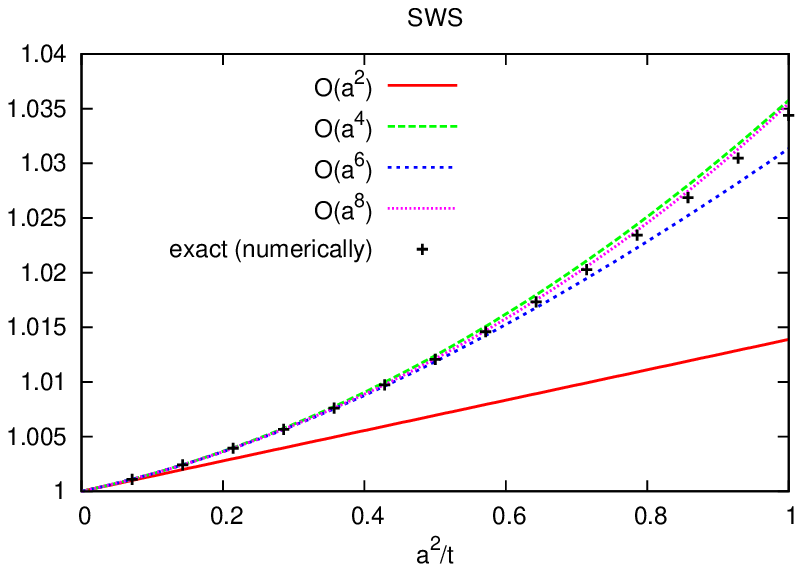}  \includegraphics[width=7.5cm]{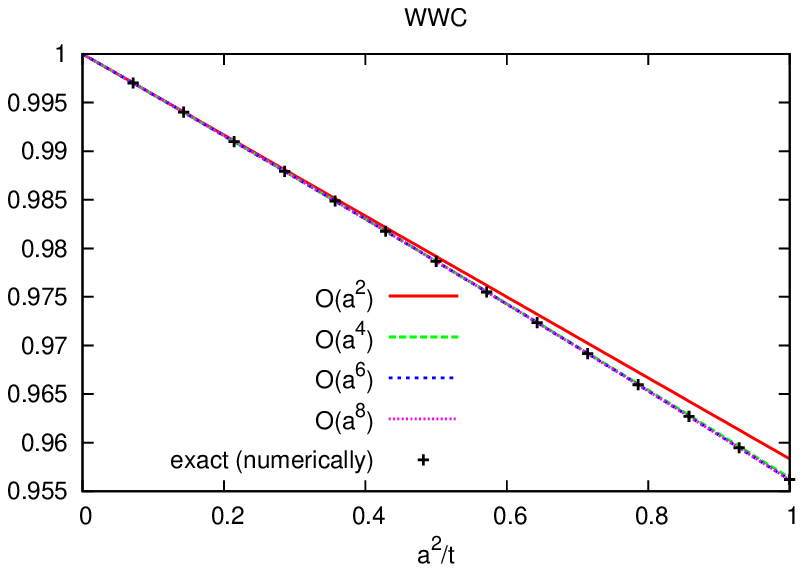} \\
\includegraphics[width=7.5cm]{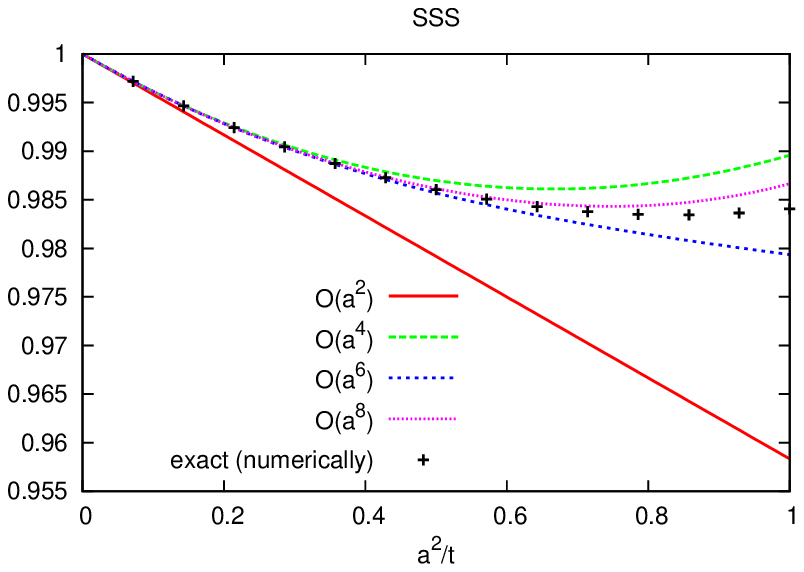}  \includegraphics[width=7.5cm]{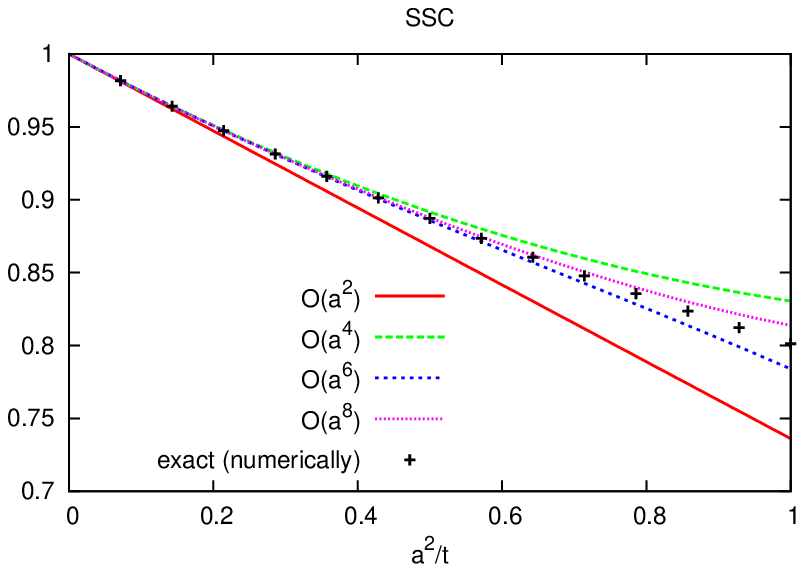} 
\end{center}
\caption{The tree-level improvement factor for four examples, the $SWS$, $WWC$, $SSS$ and $SSC$ cases, 
starting from only the $O(a^2)$ correction and including higher and
higher order terms. What is also shown is the all-order result obtained by numerically evaluating (\ref{eee}).}
\label{conv}
\end{figure}

Another cross-check we performed is the numerical evaluation of the integral (\ref{eee}). The lattice momentum integrals
are replaced by sums and if the sum is over sufficiently many terms, $N$, the integral can be approximated to arbitrary
precision by extrapolating $N\to\infty$.
This way one obtains the result to all orders in $a^2/t$, at least numerically. The obtained result can then
be compared with the correction $1 + C_2 a^2/t + C_4 a^4/t^2 + C_6 a^6/t^3 + C_8 a^8/t^4 \ldots$. 
For some combinations of Wilson plaquette ($c=0$) and tree-level improved Symanzik ($c=-1/12$) actions as 
well as the clover observable this comparison is shown in figure \ref{conv}.

Let us introduce the shorthand \sss for Symanzik improved flow, Symanzik improved gauge action 
and Symanzik improved observable, and similarly
\ssc for Symanzik improved flow, Symanzik improved gauge action and clover observable. The order is
always Flow-Action-Observable. For the frequently used cases, Wilson plaquette ($c_1 = 0$), tree-level Symanzik ($c=-1/12$)
or clover we will use abbreviations like $WWC$, $SSS$, $SSC$, etc, again with the ordering Flow-Action-Observable.

\vspace{0.5cm}
{\em \large Order $a^2$ correction}
\vspace{0.5cm}

The first correction to the continuum result is $O(a^2/t)$. Expanding the lattice momentum integral (\ref{eee}) to first
order one obtains
\bea
\label{c2sss}
C_2 = 2c_f + \frac{2}{3}c_g - \frac{2}{3}c_e + \frac{1}{8}
\eea
for the \sss case, and
\bea
\label{c2ssc}
C_2 = 2c_f + \frac{2}{3}c_g - \frac{1}{24}
\eea
for the \ssc case. Several observations are in order. 
First, clearly it is possible to choose the improvement coefficients such that these corrections are zero,
i.e. $O(a^2)$ improvement is possible in both cases. Second, $O(a^2)$ improvement does not fix all the freedom we have
in the discretization, in the \sss case we still have 2 free parameters and in the \ssc case 1 free parameter. These can
be used to improve to higher orders in the expansion.

Some of the frequently used setups in practice are the Wilson ($c=0$) and
tree-level improved Symanzik ($c=-1/12$) actions and their combinations with or without clover observable. The $O(a^2)$
term for these are listed in table \ref{c2table}. Perhaps surprisingly, the smallest cut-off effects among these frequently used
combinations is exhibited by the $SWS$ setup: $c_f = -1/12$, $c_g = 0$, $c_e = -1/12$.

In \cite{Luscher:2010iy} it was observed that the cut-off effects are much smaller with $WWC$ discretization than with $WWW$
discretization. This is consistent with the relative size of the corresponding $C_2$ coefficients in table
\ref{c2table}.

In order to simplify notation for the higher order terms let us introduce
\bea
x = 2c_f + \frac{1}{8}\;,\qquad y = c_g - \frac{1}{4}\;,\qquad z = c_g - c_e\;.
\eea
Using $(x,y,z)$ the $O(a^2)$ coefficients are simply
\bea
\label{c2sssx}
C_2 = x + \frac{2}{3} z
\eea
for the \sss case and
\bea
\label{c2sscx}
C_2 = x + \frac{2}{3} y
\eea
for the \ssc case. 

\begin{table}
\begin{center}
\bgroup
\small
\begin{tabular}{|l|c|c|c|c|c|c|}
\hline
      &       $SWS$  &        $WWC$  &        $SSS$  &         $SWW$  &           $WSW$  &          $WSC$ \\
\hline                                                                           
\hline                                                                           
$C_2$ &          1/72  &      -1/24  &        -1/24  &         -1/24  &            5/72  &          -7/72 \\
\hline                                                                          
$C_4$ &         7/320  &     -1/512  &         1/32  &          1/32  &         23/1280  &        19/2560 \\
\hline                                                                          
$C_6$ & -8539/1935360  &    -1/5120  &   -283/27648  &    -283/27648  &     2077/483840  &  -2237/1935360 \\
\hline                                                                          
$C_8$ &76819/18579456  &   -1/65536  &  3229/442368  &   3229/442368  &   16049/9289728  & 14419/74317824 \\
\hline           
\hline           
      &         $SSW$  &       $WWW$ &       $WSS$   &         $WWS$  &           $SWC$  &          $SSC$ \\
\hline                                                                     
\hline                                                                     
$C_2$ &         -7/72  &         1/8 &         1/8   &         13/72  &           -5/24  &         -19/72 \\
\hline                                                                    
$C_4$ &        35/768  &       3/128 &       3/128   &        13/384  &        167/2560  &       145/1536 \\
\hline                                                                    
$C_6$ &  -5131/276480  &     13/2048 &     13/2048   &     277/30720  &  -58033/1935360  &  -12871/276480 \\
\hline                                                                    
$C_8$ &  10957/884736  &    77/32768 &    77/32768   &     323/98304  & 457033/24772608  &  52967/1769472 \\
\hline           
\end{tabular}
\egroup
\end{center}
\caption{The $O(a^2)$, $O(a^4)$, $O(a^6)$ and $O(a^8)$ correction terms $C_{2,4,6,8}$ 
for various frequently used discretizations. W stands for Wilson ($c=0$), S for tree-level
improved Symanzik ($c=-1/12$) and C for clover. See text for more details.}
\label{c2table}
\end{table}

\vspace{0.5cm}
{\em \large Order $a^4$ correction}
\vspace{0.5cm}

Continuing the expansion in the lattice spacing to the next order we obtain the corrections to $O(a^4)$. 
Explicitly,
\bea
C_4 = \frac{57}{32} x^2 - \frac{25}{128}x + \frac{57}{40}xz + \frac{57}{80}yz + \frac{1}{8}z + \frac{41}{2048}
\eea
for the \sss case and
\bea
C_4 = \frac{57}{32} x^2 - \frac{25}{128}x + \frac{57}{40}xy + \frac{57}{80}y^2 + \frac{1}{8}y + \frac{53}{2048}
\eea
for the \ssc case.

The $C_4$ coefficients for the frequently used discretizations, Wilson ($c=0$) 
and tree-level improved Symanzik ($c = -1/12$) with or without clover observable are again listed in table
\ref{c2table}. 

In the next section we will see that $C_2$, $C_4$ and $C_6$ can all be made zero with non-conventional improvement
coefficients and even $C_8$ will be extremely small.

\vspace{0.5cm}
{\em \large Order $a^6$ correction}
\vspace{0.5cm}

Continuing the expansion in an increasingly complicated fashion one obtains
\bea
C_6 &=&
 \frac{1205}{256}x^3
-\frac{2247}{2048}x^2
+\frac{391}{3584}xz
+\frac{1205}{448}xyz
+\frac{241}{224}y^2z \nn \\
&&+\frac{6807}{17920}yz
+\frac{3615}{896}x^2z
+\frac{2191}{16384}x
+\frac{19247}{286720}z
-\frac{317}{131072}
\eea
in the \sss case, while
\bea
C_6 &=&
 \frac{1205}{256}x^3
-\frac{2247}{2048}x^2
+\frac{391}{3584}xy
+\frac{1205}{448}xy^2
+\frac{241}{224}y^3 \nn \\
&&+\frac{6807}{17920}y^2
+\frac{3615}{896}x^2y
+\frac{2559}{16384}x
+\frac{21823}{286720}y
-\frac{993}{655360}
\eea
in the \ssc case.

\vspace{0.5cm}
{\em \large Order $a^8$ correction}
\vspace{0.5cm}

The final term we calculate in this paper is the $O(a^8)$ coefficient which for the \sss case is
\bea
\label{c81}
C_8 &=&
 \frac{17181}{1024}x^4
-\frac{12361}{2048}x^3
+\frac{1909}{128}x^3z
-\frac{33}{32}x^2z
+\frac{5727}{512}x^2yz
+\frac{5727}{896}xy^2z+      \nn \\
&&+\frac{20187}{14336}xyz
+\frac{8137}{7168}y^2z
+\frac{1909}{896}y^3z
+\frac{32667}{32768}x^2
-\frac{8349}{131072}x
+\frac{4341}{229376}z+      \nn \\
&&+\frac{63121}{229376}yz
+\frac{22107}{57344}xz
+\frac{10181}{4194304}\;,
\eea
while for the \ssc case 
\bea
\label{c82}
C_8 &=&
 \frac{17181}{1024}x^4
-\frac{12361}{2048}x^3
+\frac{1909}{128}x^3y
-\frac{33}{32}x^2y
+\frac{5727}{512}x^2y^2
+\frac{5727}{896}xy^3+      \nn \\
&&+\frac{20187}{14336}xy^2
+\frac{8137}{7168}y^3
+\frac{1909}{896}y^4
+\frac{17877}{16384}x^2
-\frac{1071}{16384}x
+\frac{10999}{458752}y+     \nn \\
&&+\frac{12597}{28672}xy
+\frac{67237}{229376}y^2
+\frac{14387}{4194304}
\eea
is obtained.

As a cross-check let us look at the $WWW$ column in table \ref{c2table} corresponding to the simplest case of Wilson
plaquette flow, Wilson plaquette action and Wilson plaquette observable. In this case the exact result is easy to
compute from (\ref{eee}),
\bea
\label{www}
C(a^2/t) = 64\pi^2 t^2/a^4 \left( e^{-4t/a^2} I_0\left(4t/a^2\right) \right)^4\;,
\eea
where $I_0$ is the well-known Bessel function.
Its asymptotic expansion $t/a^2 \to \infty$ precisely matches the coefficients in table \ref{c2table}.

Figure \ref{conv} shows the correction factor for 4 cases involving the frequently used discretizations as an
illustration. More and more terms are included in the expansion all the way up to $O(a^8)$ and we also show the
numerically evaluated exact expression.

\section{Improved gradient flow}
\label{highlyimp}

Using the results from section \ref{latpert} it is clear that the 3 free parameters $c_f, c_g, c_e$ (or
equivalently $x,y,z$) can be tuned to achieve $O(a^6)$ improvement in the \sss case. In the \ssc case only 2
free parameters are available, $c_f, c_g$ (or equivalently $x,y$) and only $O(a^4)$ improvement is possible.

The 3 parameters in the \sss case will allow setting $C_{2,4,6} = 0$ and actually there are two separate sets of
solutions. Substituting both of these into $C_8$ and picking the one which is minimal, leads to the improvement
coefficients 
\bea
\label{ccxx}
c_{f} &=&-0.013993 \nn    \\
c_{g} &=& 0.052556        \\
c_{e} &=& 0.198078\;, \nn
\eea
This choice corresponds to $O(a^6)$ improvement at tree-level and even the next coefficient $C_8 = 0.0001253$ is quite small.

Similarly, in the \ssc case one may require $C_{2,4} = 0$ and obtain
\bea
\label{cx}
c_{f} &=& -0.012250      \\
c_{g} &=&  0.099250 \nn 
\eea
with $C_6 = -0.0003022$, corresponding to $O(a^4)$ improvement.

In actual simulations the gauge action used for generating the configurations is sometimes 
considered fixed and one only has
control over the measurements. In these cases one may optimize the discretization of the flow and observable 
(i.e. $c_f, c_e$) for the \sss case or the discretization of the flow only (i.e. $c_f$)  for the \ssc case. 

For example if $c_{g} = 0$, i.e. the Wilson plaquette action was used for the simulation then,
\bea
\label{wilsonc}
c_{f} &=& 0 \nn \\
c_{e} &=& 3/16
\eea
will lead to $C_{2,4} = 0$ and $C_6 = 7/20480$. As another example we
quote the case of tree-level Symanzik improved gauge action for the simulation, i.e. $c_{g} = -1/12$ in which case the
optimal improvement coefficients are
\bea
\label{symc}
c_{f} &=& 0.0388441 \nn \\
c_{e} &=& 0.2206988 
\eea
which again lead to $C_{2,4} = 0$ and $C_6 = -0.00131710$.

If the clover term is used for the observable and again $c_{g}$ is considered fixed, then setting $C_2 = 0$ is always
possible leading to $O(a^2)$ improvement. For arbitrary $c_{g}$ the optimal choice is
\bea
c_f = \frac{1}{48} - \frac{1}{3} c_g\;,
\eea
leading to $C_4 = \frac{133}{240}c_g^2 - \frac{7}{320}c_g - \frac{101}{30720}$.

\section{Scale setting}
\label{scaleset}

The tree-level coefficients $C_{2m}$ can also be used for improving scale setting by both $t_0$ \cite{Luscher:2010iy} and $w_0$
\cite{Borsanyi:2012zs}. First, let us discuss scale setting by $t_0$. In this setup the quantity $F(t)=\langle
t^2E(t)\rangle$ is set to a certain value $F_0$, usually $F_0 = 0.3$ in QCD, and the corresponding $t=t_0$ determines
the scale, $1/\sqrt{t_0} \sim \Lambda_{QCD}$. Now one may improve on the scale determination by considering the
improvement of $F(t)$ calculated in the previous sections. The improved scale setting condition is then
\bea
\label{ftexp}
\frac{F(t_{0\, imp})}{1 + \sum_{m=1}^4 C_{2m} \frac{a^{2m}}{{t_{0\, imp}}^m}} = F_0\;,
\eea
which may be solved directly for $t_{0\, imp}$ using the numerically evaluated $F(t)$. It is instructive however to
expand $t_{0\, imp}$ in the lattice spacing in order to see how small or large the effect of improvement is on this
particular quantity. 

To this end let us introduce coefficients $T_{2m}$ by
\bea
\label{texp}
t_{0\, imp} = t_0 \left( 1 + \sum_{m=1}^4 T_{2m} \frac{a^{2m}}{t_0^m} \right)\;.
\eea
Clearly, $t_{0\, imp}$ needs to be expanded in $a^2/t_0$ according to (\ref{texp}) at two instances in (\ref{ftexp}), first in
the argument of $F(t_{0\, imp})$ and second, in the denominator. The expansion then leads to the determination of the
coefficients $T_{2m}$. In order to simplify notation let us introduce for the derivatives,
\bea
F^\prime_0 &=& t \left.\frac{d}{dt}F(t)\right|_{t=t_0} \nn \\
F^{\prime\prime}_0 &=& t^2 \left.\frac{d^2}{dt^2}F(t)\right|_{t=t_0} \\
F^{\prime\prime\prime}_0 &=& t^3 \left.\frac{d^3}{dt^3}F(t)\right|_{t=t_0}\;. \nn 
\eea
Using these the sought after improvement coefficients of $t_{0\,imp}$ are,
\bea
\label{t0res}
T_2 &=& C_2 \frac{F_0}{F^\prime_0} \\
T_4 &=& C_4 \frac{F_0}{F^\prime_0} - C_2^2 \frac{F_0^2}{{F^\prime_0}^3}\left( F^\prime_0 + \frac{1}{2}F^{\prime\prime}_0
\right) \nn \\
T_6 &=& C_6 \frac{F_0}{F^\prime_0} - C_2C_4 \frac{F_0^2}{{F^\prime_0}^3} \left( 3 F^\prime_0 + F^{\prime\prime}_0 \right) + 
C_2^3 \frac{F_0^3}{{F^\prime_0}^5} 
\left( 2 {F^\prime_0}^2 + \frac{3}{2} F^\prime_0 F^{\prime\prime}_0 + \frac{1}{3} {F^{\prime\prime}_0}^2 - \frac{1}{6}
F^{\prime\prime\prime}_0 F^\prime_0 \right)\;. \nn
\eea
The term $T_8$ is straightforward to calculate as well but will not be quoted here as it is quite lengthy.
Note that in QCD the shape of $F(t)$ is roughly linear in $t$ for $t \sim t_0$ hence 
$F^{\prime\prime}_0$ and all further derivatives will be small.

A useful variant of scale setting by $t_0$ was introduced in \cite{Borsanyi:2012zs} called $w_0$. In this setup the
logarithmic derivative of $F(t)$ is set to a prescribed value $F^\prime_0$, again usually $F^\prime_0 = 0.3$ in QCD,
and the corresponding $t=w_0^2$ value is used as scale, $w_0 \sim 1/\Lambda_{QCD}$,
\bea
\left. t\frac{d}{dt}F(t) \right|_{t=w_0^2} = F^\prime_0\;.
\eea
In a completely analogous way as done for $t_0$ one may introduce the improved scale $w_{0\;imp}$ by
\bea
\label{w0}
\left[ t\frac{d}{dt}\frac{F(t)}{1 + \sum_{m=1}^4 C_{2m} \frac{a^{2m}}{t^m} } \right]_{t=w_{0\,imp}^2} = F^\prime_0\;,
\eea
which again can be solved directly. It is again instructive to expand $w_{0\,imp}^2$ in the lattice spacing,
\bea
w_{0\,imp}^2 = w_0^2 \left( 1 + \sum_{m=1}^4 W_{2m} \frac{a^{2m}}{w_0^{2m}} \right)\;,
\eea
using another set of coefficients $W_{2m}$ which can be determined also analogously.
When (\ref{w0}) is expanded one is lead to
\bea
W_2 &=& C_2 \frac{F^\prime_0 - F_0 }{F^{\prime\prime}_0 + F^{\prime}_0}\;,
\eea
where of course the index $0$ refers to evaluating the function $F$ and its various derivatives at $t=w_0^2$.
The further terms $W_{4,6,8}$ can also be easily calculated but are rather lengthy.

\section{Finite volume}
\label{finitevolume}

So far the calculations were performed on an infinite lattice. In \cite{Fodor:2012td} the corresponding calculations in
the continuum but finite volume were performed. In the present section we combine the two approaches and obtain results
at finite lattice spacing and finite volume.

The gauge field is assumed to be periodic in all 4 directions and the result will have two distinct contributions, one
coming from the zero modes and a second one from the non-zero modes.
It can be shown that the contribution of the zero mode is the same in the
continuum as at finite lattice spacing \cite{Coste:1985mn}. This is intuitively clear 
because the zero mode is a constant and the
4-torus can just as well be considered a single point. In this case of course discretization effects cannot come into
play. Hence only the contribution of the non-zero modes can be lattice spacing dependent. This contribution can
easily be evaluated numerically by replacing the integral in our formula (\ref{eee}) by a discrete lattice sum over
non-zero 4-momenta.

In this way we obtain the lattice spacing dependence of the finite volume correction factor $\delta(\sqrt{8t}/L)$ of
\cite{Fodor:2012td} where the ratio $\sqrt{8t}/L$ was called $c$ but in order not to introduce confusion with the
improvement coefficients, $c$ will not be used for the ratio here.
Equivalently, we obtain the finite volume dependence of the finite lattice spacing correction factor
$C(a^2/t)$ of the present work,
\bea
C(a^2/t,\sqrt{8t}/L) = 1 + \delta(\sqrt{8t}/L, a/L)\;.
\eea
Specifically, using the formula (\ref{eee}) and the finite volume results from \cite{Fodor:2012td} we obtain at
finite lattice spacing and finite volume and leading order in the coupling,
\bea
\label{catv}
\langle t^2 E(t) \rangle &=& g^2 \frac{3(N^2-1)}{128\pi^2} C(a^2/t,\sqrt{8t}/L) \\
C(a^2/t,\sqrt{8t}/L) &=& \frac{128\pi^2 t^2}{3L^4} + \frac{64\pi^2 t^2}{3L^4} 
\sum_{n_\mu = 0,\; n^2\neq 0}^{L/a-1} \tr\, \left( e^{-t\left({\cal S}^f + {\cal G}\right)} ({\cal S}^g + {\cal G})^{-1}
e^{-t\left({\cal S}^f + {\cal G}\right)} {\cal S}^e \right)\;, \nn
\eea
where again the first term comes from the zero modes and is identical to the continuum result and $p_\mu = 2\pi n_\mu
/L$ with a non-zero integer 4-vector $n_\mu$. This expression can
easily be evaluated numerically for any choice of discretizations.

\begin{figure}
\begin{center}
\includegraphics[width=7.5cm]{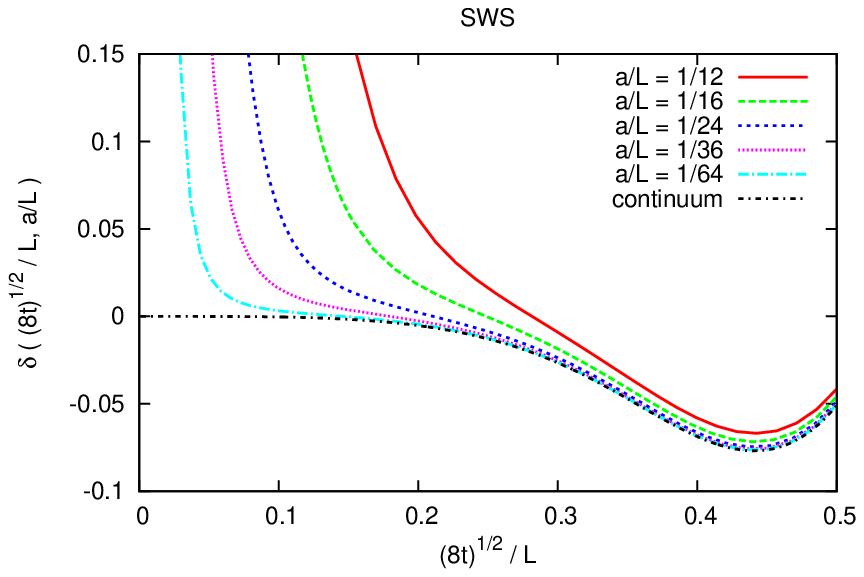}  \includegraphics[width=7.5cm]{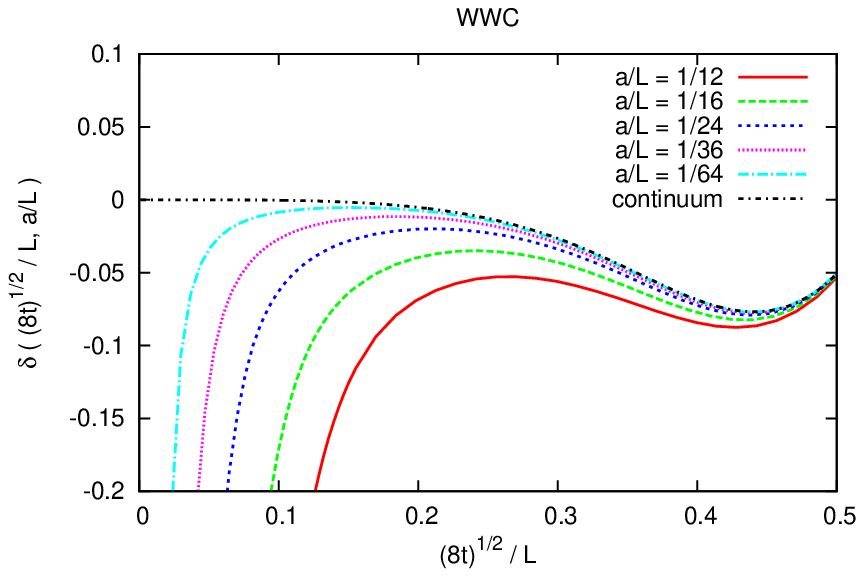} \\
\includegraphics[width=7.5cm]{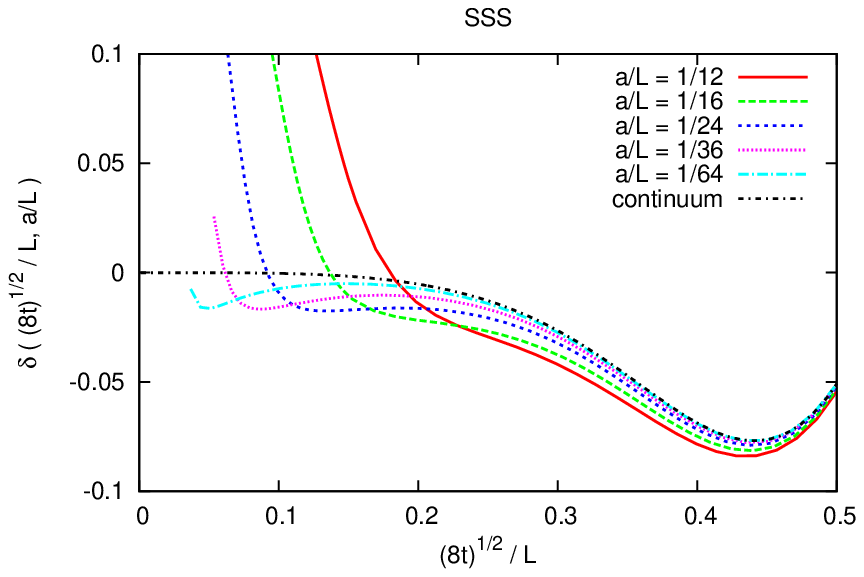}  \includegraphics[width=7.5cm]{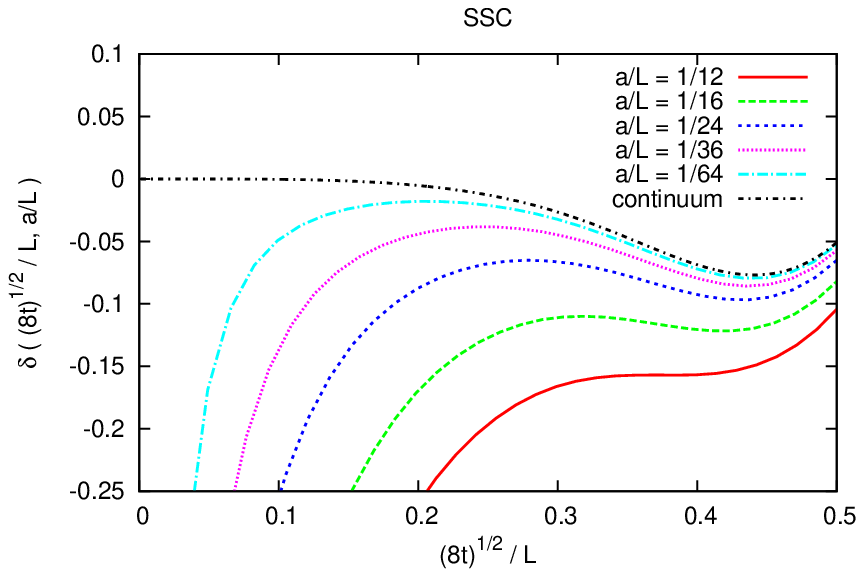} 
\end{center}
\caption{The tree-level finite volume and finite lattice spacing correction factors
$\delta(\sqrt{8t}/L, a/L) = C(a^2/t,\sqrt{8t}/L) - 1$
for four examples, the $SWS$, $WWC$, $SSS$ and $SSC$ cases as a function of $\sqrt{8t}/L$ at various lattice spacings.
The continuum result is from \cite{Fodor:2012td}.}
\label{delta}
\end{figure}

For illustration we plot $\delta(\sqrt{8t}/L,a/L) = C(a^2/t,\sqrt{8t}/L) - 1$ for four examples at various lattice volumes as a
function of $\sqrt{8t}/L$ on figure \ref{delta}. 
We also show the continuum result $\delta(\sqrt{8t}/L)$ from \cite{Fodor:2012td} for comparison.

\section{Numerical test}
\label{numericaltest}

In order to test the numerical usefulness of our tree-level formulae we will consider the running coupling of $N_f = 4$
flavors \cite{Fodor:2012td}. For all details of the simulations we refer to the original work \cite{Fodor:2012td}, here we
simply quote two examples of continuum extrapolations that were performed there. At these and all the other 
renormalized couplings we computed the discrete $\beta$-function corresponding to a scale change of
$s=3/2$ at various lattice spacings and performed continuum extrapolations. 

Since the setup in \cite{Fodor:2012td} is a step-scaling approach to the calculation of the $\beta$-function on a
periodic 4-torus we need to use the finite volume, finite lattice spacing factor computed in the previous section. 

For our numerical test the renormalized coupling in \cite{Fodor:2012td} is tree-level improved by dividing by the tree-level
expression $C(a^2/t,\sqrt{8t}/L)$ from (\ref{catv}) in the $SSC$ setup, i.e. tree-level Symanzik
improved ($c=-1/12$) gauge action and flow with clover type observable with $\sqrt{8t}/L = 3/10$, 
which was the setup used in the simulation. As it can be seen from figure \ref{delta} the continuum finite volume
correction factor at $\sqrt{8t}/L = 3/10$ is around 3\% while the finite lattice spacing dependence can be 12\% away
from it for the smaller lattices while only about 1\% away from it for the larger $L/a = 36$ lattices.

\begin{figure}
\begin{center}
\includegraphics[width=7.5cm]{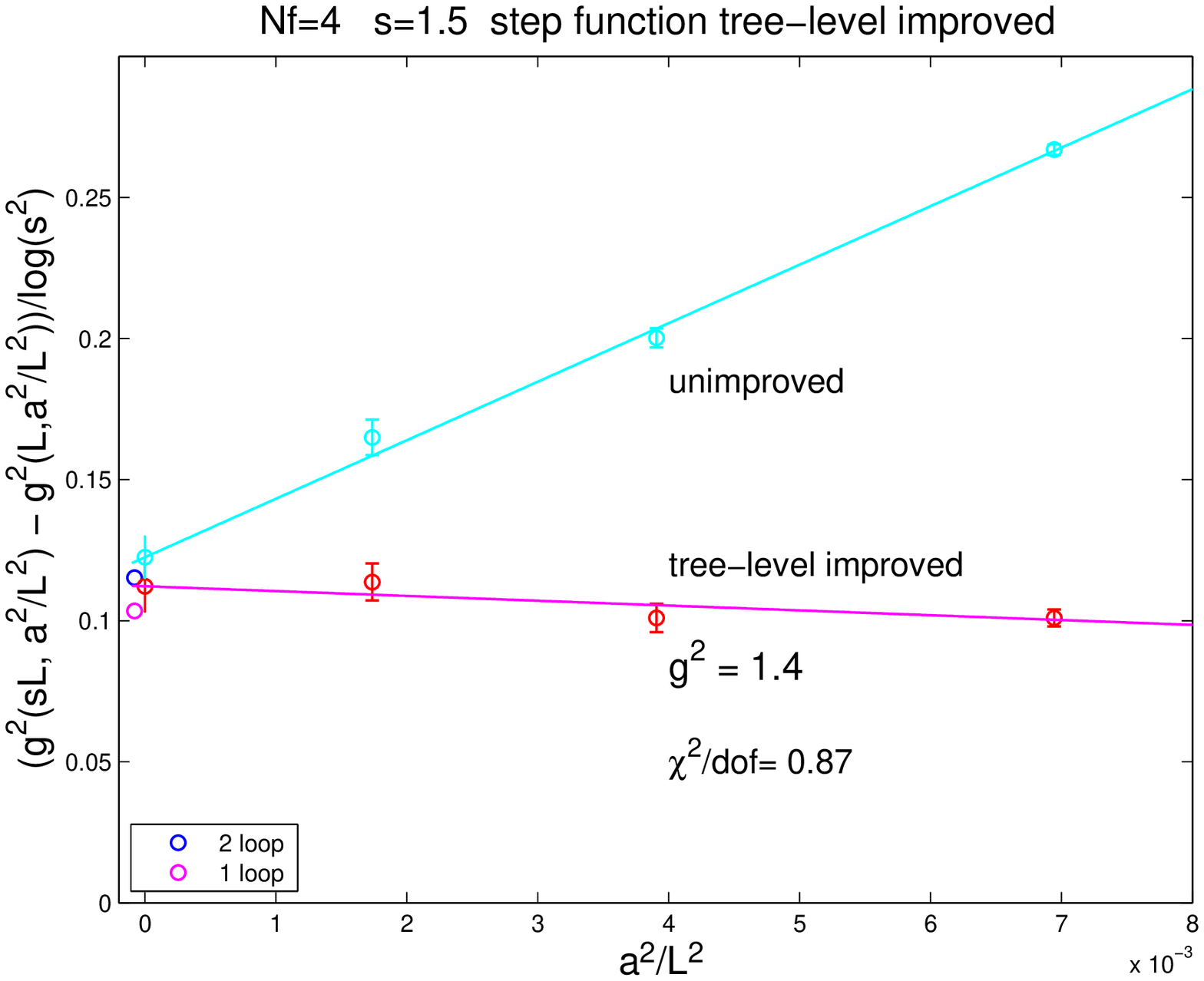} \includegraphics[width=7.5cm]{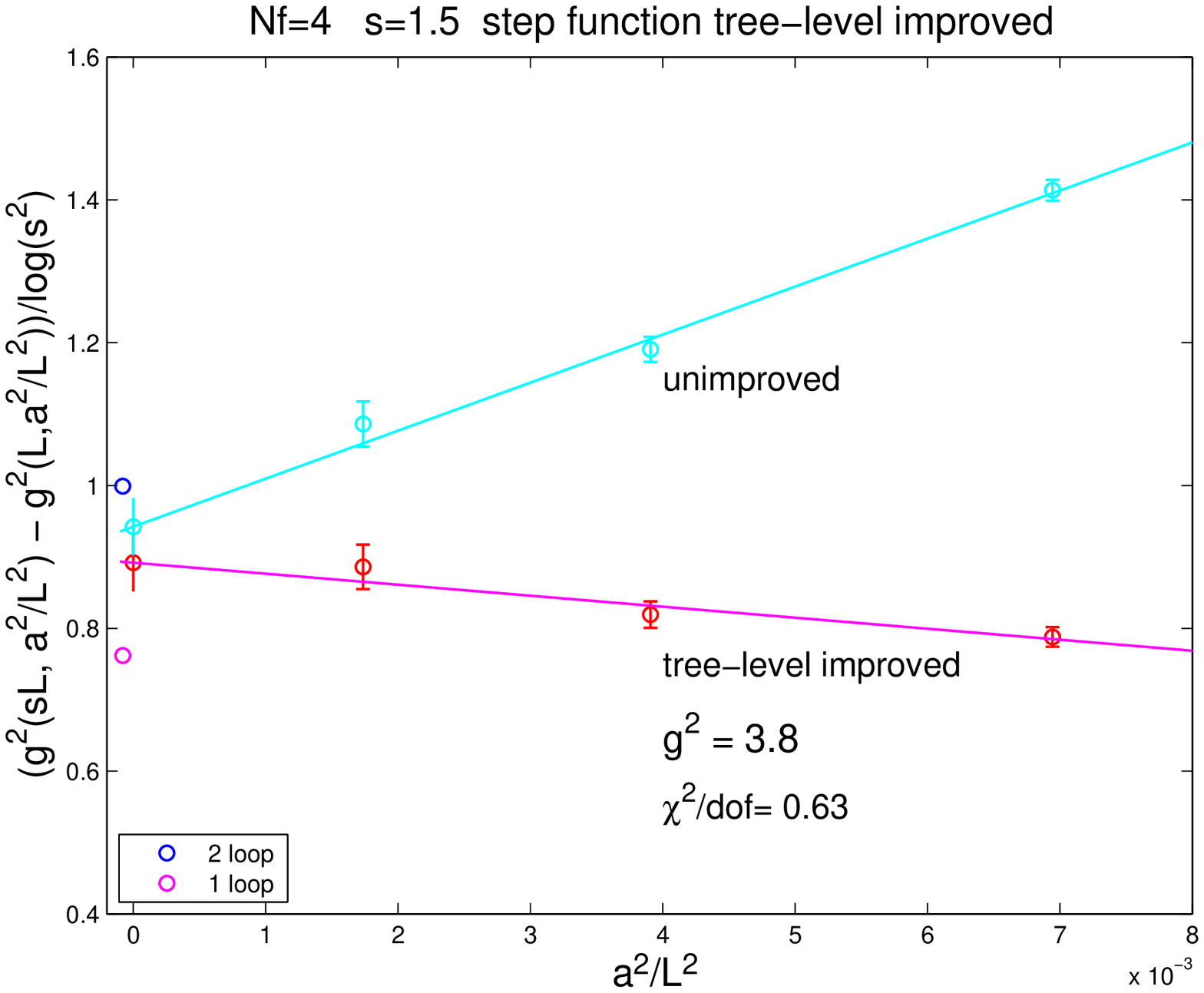}
\end{center}
\caption{Continuum extrapolations of the discrete $\beta$-function for two selected $g^2$ values $1.4$ (left) and
$3.8$ (right) for $N_f = 4$ flavors with and without tree-level improvement. The data is from \cite{Fodor:2012td}.}
\label{beta}
\end{figure}

\begin{figure}
\begin{center}
\includegraphics[width=7.5cm]{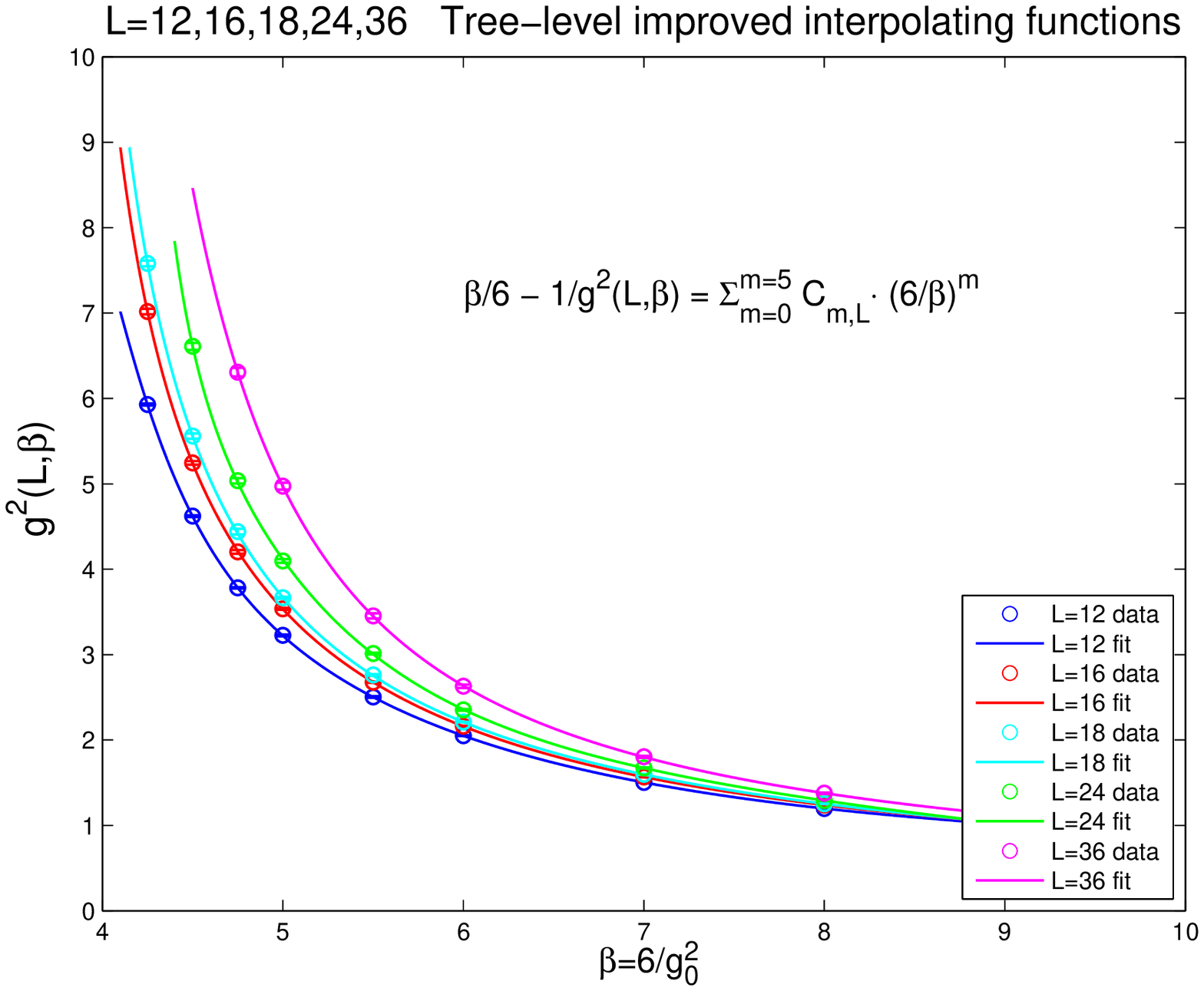} \includegraphics[width=7.5cm]{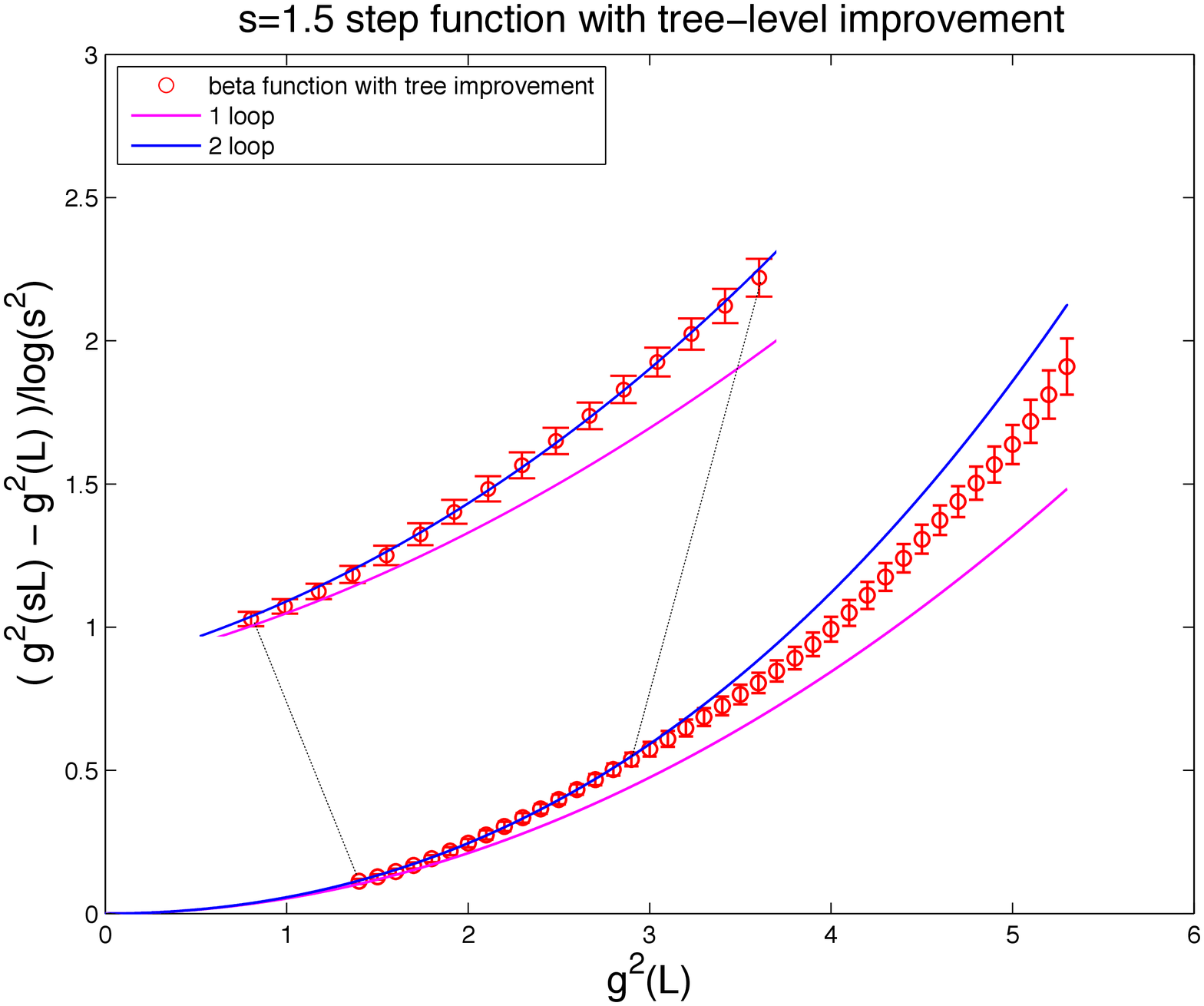}
\end{center}
\caption{Parametrization of the tree-level improved renormalized coupling at fixed 
lattice sizes as a function of $\beta$ (left) and 
continuum extrapolated discrete $\beta$-function of $N_f = 4$ flavors with a magnified section for better
visibility (right). The data is from \cite{Fodor:2012td}.}
\label{beta2}
\end{figure}

The resulting tree-level improved continuum extrapolations are shown in figure \ref{beta}. Clearly, the slope of the
extrapolation is greatly reduced by the improvement.
The final error on the improved extrapolation is somewhat smaller than the unimproved version as expected.

Figure \ref{beta2} shows the interpolation of the tree-level improved renormalized couplings on various
lattice volumes as a function of $\beta$ and also the final continuum extrapolation for the discrete $\beta$-function
for the full range of the renormalized couplings together with the 1-loop and 2-loop results. It should be noted though
that the displayed error on the continuum result for the discrete $\beta$-function (figure \ref{beta2}, right) only
includes statistical errors. Systematic effects so far have not been estimated. 

\section{Conclusion and outlook}
\label{conc}

The tree-level results obtained in this paper are useful for two reasons. First, if any simulation is performed with
an arbitrary discretization the tree-level results can be used to tree-level improve the non-perturbative results by
the corresponding tree-level expression. The necessary expressions were given for both finite and infinite volume
setups. Such a tree-level improvement of course will not affect the continuum
extrapolated result but will reduce the size of cut-off effects. Finite volume tree-level improvement was 
demonstrated to be useful for the calculation of the running coupling and the corresponding $\beta$-function.

Second, the tree-level analysis shows how to choose
optimal discretizations for all three ingredients of the calculation: the gradient flow, the gauge action and the
observable $E$, leading to $O(a^6)$ improvement. If the discretization of the dynamical gauge action is considered
fixed, $O(a^4)$ improvement can be achieved.

A natural next step in the improvement program would be to calculate the 1-loop terms at finite lattice spacing which is
beyond the scope of the present paper. 

\acknowledgments

This work was supported by the DOE under grant DE-SC0009919, by the EU Framework Programme 7 grant (FP7/2007-2013)/ERC No 208740, 
by the Deutsche Forschungsgemeinschaft SFB-TR 55 and by the NSF under grants 0704171, 0970137 and 1318220
and by OTKA under the grant OTKA-NF-104034. KH wishes to thank the Institute for Theoretical Physics and the Albert
Einstein Center for Fundamental Physics at Bern University for their
support. DN would like to thank Stefan Sint and Stefano Capitani for very helpful discussions.

\end{document}